\documentclass{amsart}
\usepackage{amsfonts}

\begin{document}
\title[Molecular dynamic simulation]{Molecular dynamic simulation of the
fluid concentration fluctuations in two dimensional porous medium}
\author{M.N. Ovchinnikov}
\address{Physics Department, Kazan State University, Kazan, Russia}
\email{Marat.Ovchinnikov@ksu.ru}
\date{\today}

\begin{abstract}
The filtration of fluid in 2D porous medium is simulated by the molecular
dynamics technique. The high concentration of fluid is created at the
initial point in time and the number of fluid particles is investigated in
all porous. The concentration of particles pores and parameters of
interaction potentials is varied. In all cases the particles concentration
decay is well described by exponential functions. The decreasing of the
interpore channels leads to the increasing of the relaxation time and does
not depend on the system size. This phenomenon is observed in real
filtration processes.
\end{abstract}

\maketitle

In recent years, much attention has been devoted to the problem of
description the non-stationary filtration in porous mezo and macro
structures \cite{1,2}. In this paper the evolution of density fluctuations
is considered by the molecular dynamics simulation \cite{3,4}. The density
of particles is related by fluid or gas pressure in modeling 2D porous media.

The particles of solid state porous matrix interacting each other by means
of Lennard and Jones potential

\begin{equation}
U\left( r\right) =4\varepsilon \left[ \left( \frac{\sigma }{r}\right)
^{12}-\left( \frac{\sigma }{r}\right) ^{6}\right]  \label{eq1}
\end{equation}

and particles of fluid interacting each other and with particles of porous
matrix via potential

\begin{equation}
U=\alpha \left( r-\rho \right) ^{2}  \label{eq2}
\end{equation}

where r -- the distances between particles, $\varepsilon $=1, $\sigma $=2$%
^{-1/6}$ , parameter $\alpha $ varies from 0 to 100, $\rho $ varies from 0
to 1. The mass of all particles was equal to unity (m=1).

2D porous system is constituted of the square lattice with porous cells. The
particles of solid state matrix embedded at sites in the lattice. In the
borders of the system under considerate are positioned the fixed particles.
The parameters $\alpha $ and $\rho $ are adopted so that matrix particles
hesitate near the equilibrium position. The matrix particles simulate the
compressibility and elastic waves propagation in porous medium. The average
distance between them was equal to unity. Into a pores was and transferred a
fluid particles. The potentials (\ref{eq1})\ and (\ref{eq2})\ imitated the
situation when the fluid particles move in pores and inter pores channels
and have a translation agility.

In our computer simulation the considerable fluctuation of the particles
number produced in pore with index (1,1). Other pores have more less number
of particles in this time moment. Then the processes of redistribution of
fluid particles between pores begin.

First the number of particles increase in pores (1,2), (2,1) and (2,2) then
in (1,3), (3,1), (3,2) and etc. At last the quasi equilibrium state is
established in all pores. Then we vary the size of systems and the
concentration of particles in pore (1,1).

In Fig.1 is shown the time dependence of particles number in pore (1,1). The
first curve is associated the condition when the concentration in pore (1,1)
was more than in others 3 times, the second curve 15 times and the third
curve 81 times ones. It is easily to seen that in all cases the kind of
decay of the number of particles in pore (1,1) is similar and may be
described by exponential type function

\begin{equation}
N\left( t\right) =N_{0}\exp \left( -\frac{t}{\tau _{char}}\right)
\label{eq3}
\end{equation}

where $N(t)$ -- the number of particles in current time moment, $N_{0}$ --
in initial time moment ones, $t_{char}$ - characteristic time of decay the
elevated density of the particles number in pore. The number of particles is
normalized to the maximal number. In Fig.1 are shown the exponential
approximations by solid lines also. $t_{char}$ in our experiments varies
from 1 to 100.

In other pores are observed the increasing of particles density. Next the
density is reduced as the initial indignation penetrates deep into system.
This process is shown in Fig.2 for the pores (1,2), (2,2) and is compared
with pore (1,1). We see that the maximum of particles is observed in pore
(1,2) at $t$=2.5 and in pore (2,2) at $t$=5. Then the perturbations transfer
into pores (1,3), (2,3), (3,3) etc.

It is an interesting to know what will be happen if we have narrowed the
interpore channels. We may do it varying the potential parameters and
increasing the repelling forces between matrix particles and fluid particles
ones. This procedure will be imitate the narrowing of the inter channels
distances. Then the permeability of porous medium will reduce. Let us denote
as $\rho $ the characteristic radius of the repelling force in potential (%
\ref{eq2}). This quantity may be named as effective width of the interpore
leaking field. We expect that characteristic time $t_{char}$ will be
increasing when the $\rho $ rise. Such is indeed the case in our
experiments. In Fig.3 is demonstrated the dependence of $t_{char}$ from $%
\rho /r_{\max }$ for the pore (1,1). Here $r_{\max }$ is the one half of the
distances between the nearest neighbor matrix particles and $\rho /r_{\max }$
may be named as relatively width of the field interpore leaking. All the
other parameters in this kind of computer simulation are fixed. We see with
growing of the $\rho /r_{\max }$ the filtration resistance becomes essential
and the $t_{char}$ considerably increase. This fact we may interpret as the
decreasing the piezoconductivity when the permeability decreases also.
Experiments shown that the size of system does not influence essentially on
the value of tchar.

\section*{Conclusion}

In our computer experiments the decay of the concentration of the particles
in separate pore is well described by exponents. This leads to the validity
of relaxation theories \cite{1}. The time of relaxation does not depend on
the size of the system. The increasing of the repelling forces between the
fluid and matrix particles tends to increase the time of relaxation of fluid
particles in pores, what corresponds to the growing of filtration resistance.

\newpage

\section*{Figure captions}

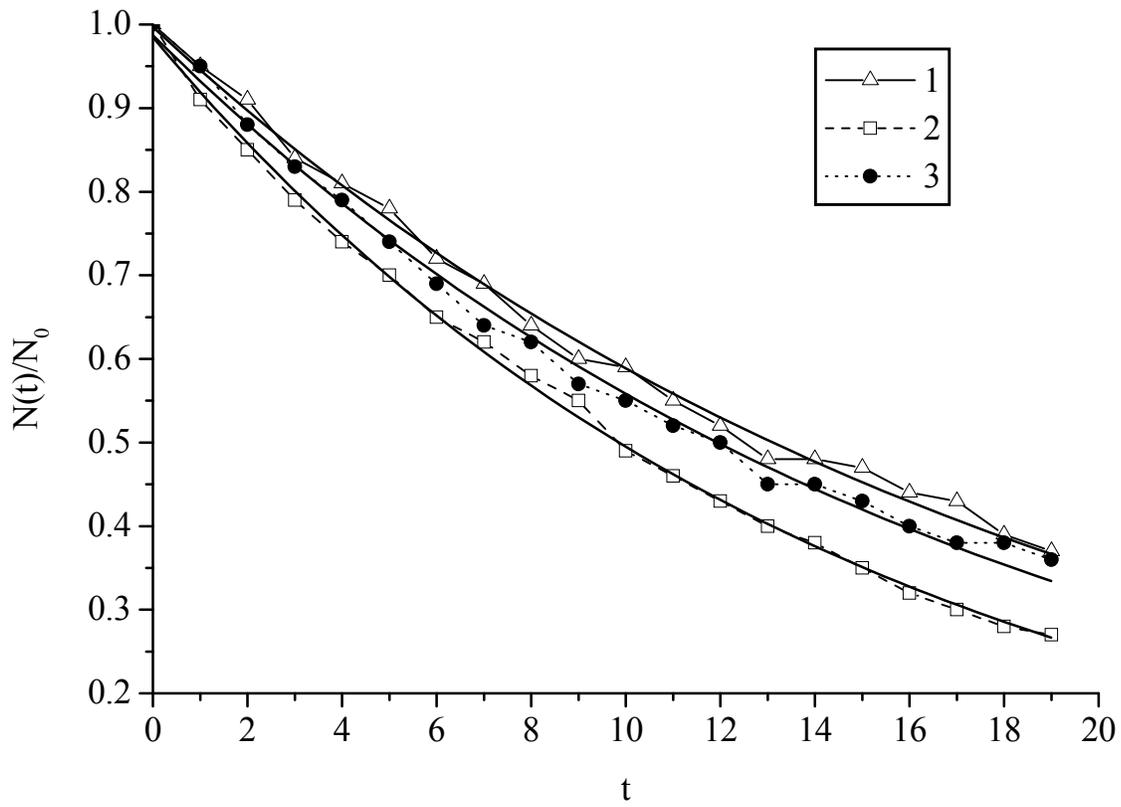
Fig.1. The time dependence of the particles number in pore (1,1). The
density of the particles in pore (1,1) more than in others 3 times (curve
1), 15 times (curve 2), 81 times (curve 3).

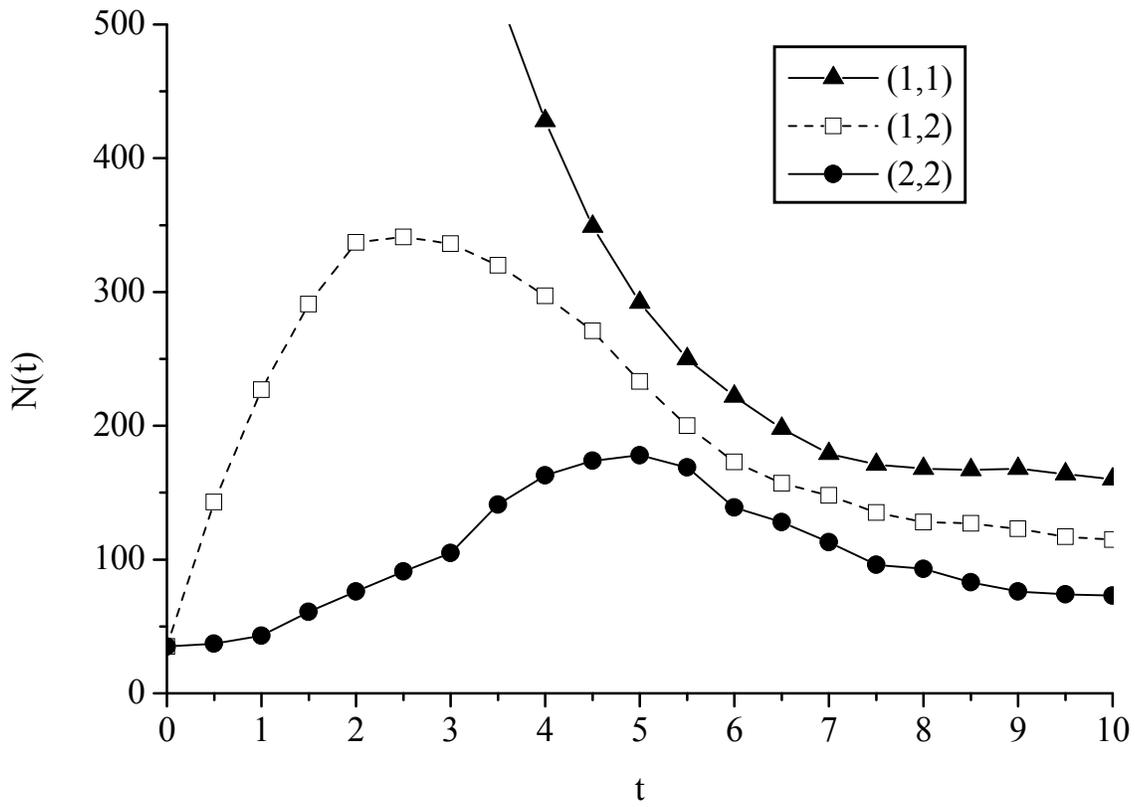
Fig.2. The time dependence of the number of particles in pores (1,2), (2,2)
and (1,1).

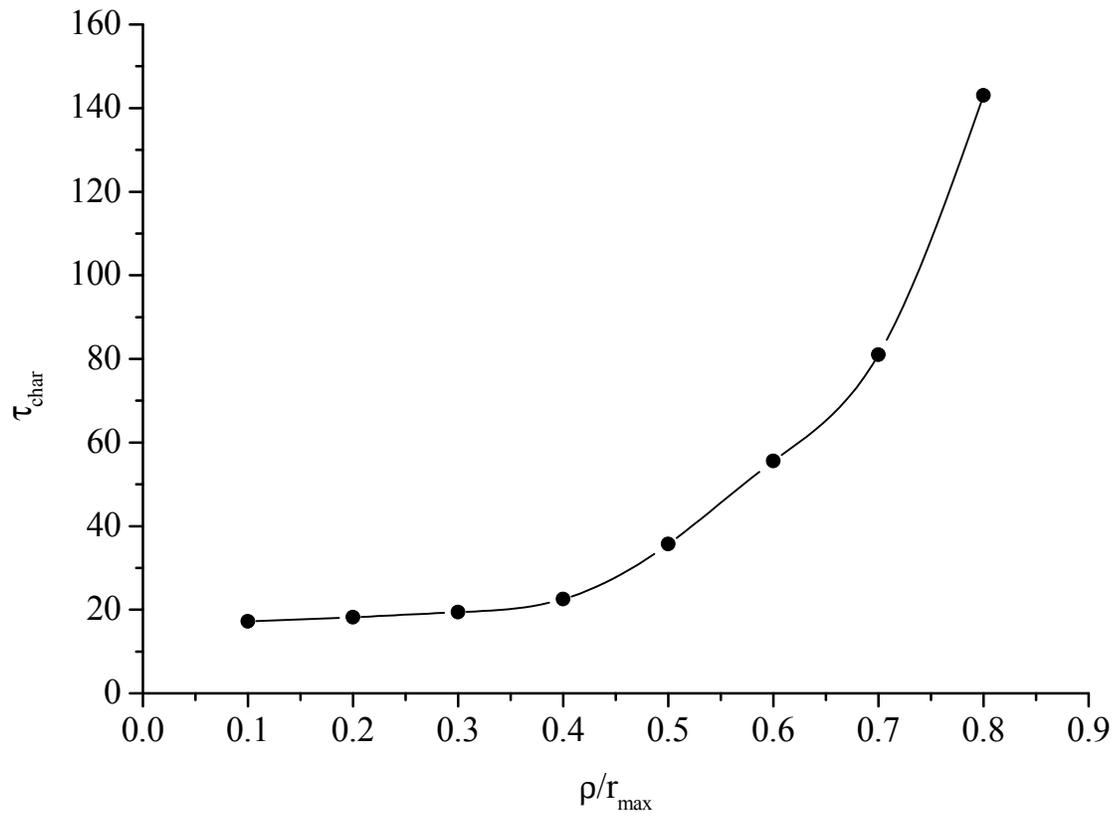
Fig.3. The dependence of the characteristic time of particles density decays
in pore (1,1) from effective width of the leaking field.

\end{document}